\documentclass[runningheads]{llncs}
\usepackage{graphicx}
\usepackage{fancyvrb}
\usepackage{hyperref}
\usepackage{cleveref}
\usepackage{syntax}
\usepackage{inputenc}
\begin{document}
\title{Bitemporal Property Graphs to Organize Evolving Systems}
\subtitle{Towards the development of a graph model, database, and query language to represent, store, and query bitemporal graphs}
%
%\titlerunning{Abbreviated paper title}
% If the paper title is too long for the running head, you can set
% an abbreviated paper title here
%
\author{Christopher Rost\inst{1,2}\orcidID{0000-0003-4217-9312} \and
Philip Fritzsche\inst{1,4} \and
Lucas Schons\inst{1,2} \and
Maximilian Zimmer\inst{1} \and
Dieter Gawlick\inst{3}\orcidID{0000-0002-7882-0565} \and\\
Erhard Rahm\inst{1,2}\orcidID{0000-0002-2665-1114}}
\authorrunning{C. Rost et al.}
% First names are abbreviated in the running head.
% If there are more than two authors, 'et al.' is used.
%
\institute{University of Leipzig, Augustusplatz 10, 04109 Leipzig, Germany \and
ScaDS.AI Dresden/Leipzig, Humboldtstraße 25, 04105 Leipzig, Germany
\email{\{rost,rahm\}@informatik.uni-leipzig.de} \and
Oracle Server Technologies, 500 Oracle Parkway, CA 94065, USA\\
\email{dieter.gawlick@oracle.com} \and
Oracle Labs Zurich, Hardstrasse 201, 8005 Zürich, Switzerland\\
\email{philip.fritzsche@oracle.com}}
\maketitle              % typeset the header of the contribution
\begin{abstract}
This work is a summarized view on the results of a one-year cooperation between Oracle Corp. and the University of Leipzig. The goal was to research the organization of relationships within multi-dimensional time-series data, such as sensor data from the IoT area. We showed in this project that temporal property graphs with some extensions are a prime candidate for this organizational task that combines the strengths of both data models (graph and time-series). The outcome of the cooperation includes four achievements: (1) a bitemporal property graph model, (2) a temporal graph query language, (3) a conception of continuous event detection, and (4) a prototype of a bitemporal graph database that supports the model, language and event detection.

\end{abstract}
\section{Introduction}
The main goal of the project was to research the suitability of temporal property graphs for the organization of multi-dimensional time series data, such as sensor data from the IoT area. Even though this data has a very high information content in itself, the used data structure offers no possibility of depicting or describing the relationships between entities, e. g., those producing time-series. 

\begin{figure}[h]
    \centering
    \includegraphics[width=0.9\textwidth]{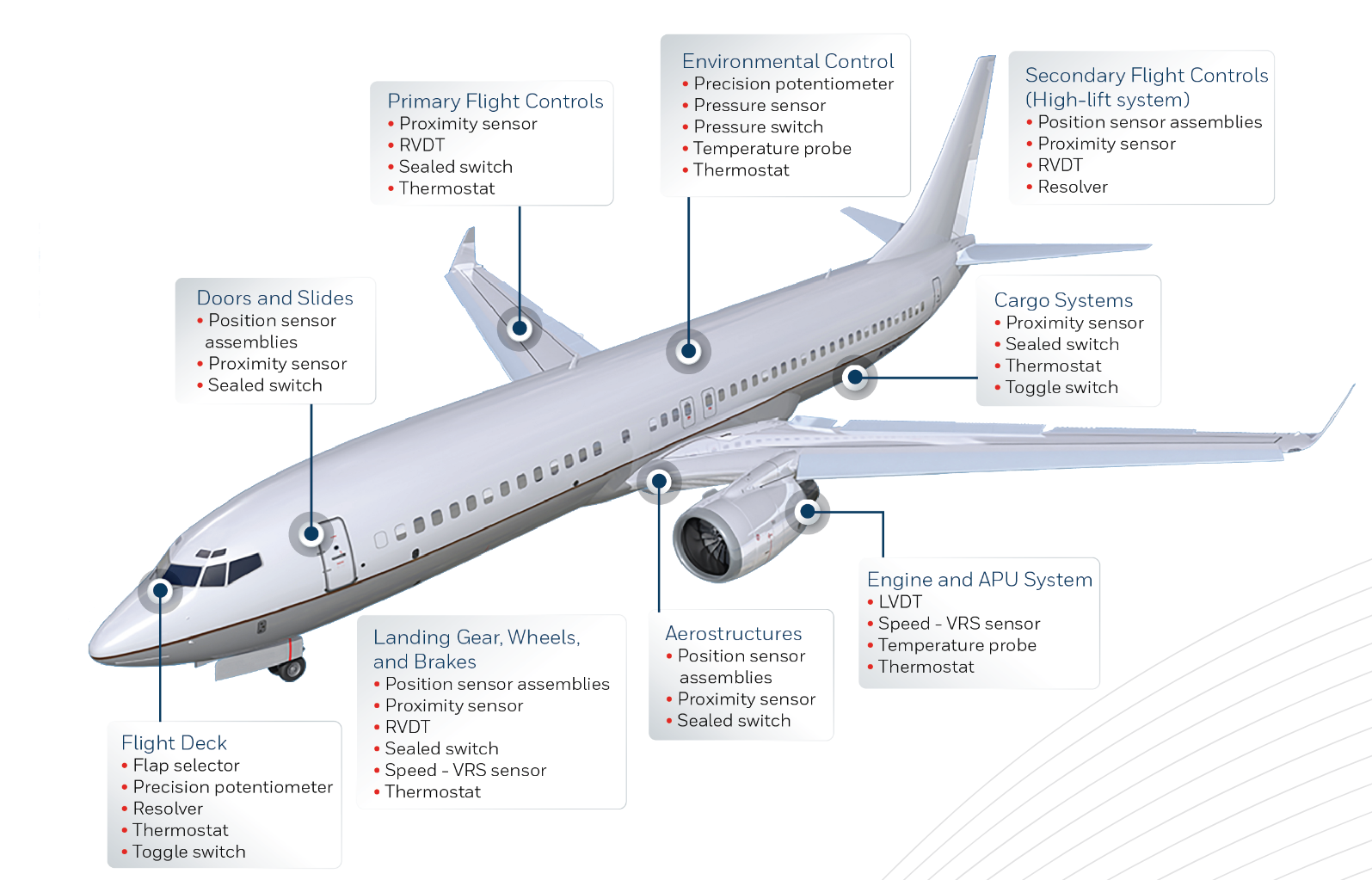}
    \caption{Sensors of an aeroplane~\cite{plane}.}
    \label{fig:plane}
\end{figure}

For example, an aircraft, like the one in Fig.~\ref{fig:plane}, is a complex system made up of a large number of bigger and smaller components, many of which are equipped with a wide variety of sensors that continuously capture data to determine its current status. One component (also denoted as an asset) is often also a complex system of smaller components, e.g., the airplane turbine, most of them again equipped with sensors.

Imagine that each sensor in such a complex system delivers a time series of values, e.g. temperatures, rotational speeds, centrifugal forces, accelerations, etc. To model the relationships between sensors and components itself; as well as sensors and components among each other, a data structure is required that can model such a complex, heterogeneous network and allow structural and content changes to the network and store exactly when these changes took place. Having such a network (or graph), one can query for these relationships and find correlations that might be hidden by looking at the time series isolated from each other. 

We figured out in this project that temporal property graphs are the prime candidate for this organizational task that combines strengths of both data models (graph and time series). Several requirements emerged for this technology combination, including (1) the need for a rich graph data model with full auditing of the graph's evolution - in the observed real world as well as in the graph database, (2) the need for a declarative query language to match patterns in a changing graph and (3) the need for an event detection mechanism through a continuous evaluation of registered patterns against graph changes.

Within the project lifetime, we achieved the following four contributions:
\begin{enumerate}
    \item TPGM$^+$: A bitemporal property graph model supporting property changes
    \item T-PGQL: A temporal graph query language
    \item CGN: Event detection through Continuous Graph Notifications
    \item BiTeGra: A bitemporal graph database based on RDBMS
\end{enumerate}

After a brief overview of related work (Section~\ref{sec:relatedwork}), we will begin by introducing a new bitemporal property graph model which supports changing properties in Section~\ref{sec:tpgm}. We further give an introduction to T-PGQL, a temporal graph query language based on Oracle's PGQL, in Section~\ref{sec:tpgql}. The Continuous Graph Notification (CGN), a technology to continuously evaluate registered T-PGQL queries for event detection purposes, is introduced in Section~\ref{sec:cgn}. In Section~\ref{sec:bitegra}, a prototype of a bitemporal graph database based on a relational database is presented, which supports the developed graph data model, the query language and event detection engine. We sum up the work and give an outlook to ongoing work in Section~\ref{sec:summary}.

\section{Related work}
\label{sec:relatedwork}

A prominent graph model is the Property Graph Model (PGM)~\cite{rodriguez2010propertygraph,angles18pgm}, which defines a directed multigraph, where vertices and edges can have an arbitrary number of \textit{properties} that are typically represented as key-value pairs. Since this model has no native maintenance of a graph's evolution, several other data models for temporal graphs exist in the literature~\cite{holme2012temporal,kostakos2009temporal,casteigts2012time,steer2020raphtory}, including duration-based graphs~\cite{wu2014path}, interval-based graphs~\cite{rost2019analyzing,rost2021distributed,campos2016towards} and snapshot graphs~\cite{khurana2013efficient}. The models differ also in other aspects, e.g., supported time-dimensions or possible updates, so that there is not yet a consensus about the most promising approach~\cite{rost2021distributed}.

Well-known declarative query languages for property graphs are: Cypher~\cite{cypher,francis2018cypher}, mainly supported by the graph database engine Neo4j, Gremlin~\cite{rodriguez2015gremlin}, part of the Apache TinkerPop graph computing framework, and PGQL~\cite{van2016pgql} which is implemented in Oracle Spatial and Graph~\cite{oracleGraph} and PGX~\cite{sevenich2016using}. These languages support date and time formats for vertex and edge properties or offer a type for expressing a duration, but they are based on the static property graph model and therefore do not support temporal property graphs, where the time is a dimension of the data model. Although temporal language extensions were introduced for relational databases a decade ago~\cite{johnston2014bitemporal,kulkarni2012temporal,jensen99}, for temporal graphs, there are just a few concepts of query languages that support the additional time dimension(s) and allow querying past graph states or formulate path patterns with chronological order. Temporal-GDL~\cite{rost2021distributed} and  T-GQL~\cite{debrouvier2021model} are two examples of temporal graph query languages that were recently introduced in the literature. 

According to the \textit{continuous} querying of graph data, most concepts are based on the streaming model, i. e., relationships (and in some models entities) are represented as an event stream that can be analyzed in (near) real-time. Frameworks supporting streaming graphs maintain a dynamically changing graph dataset under a series of updates and queries to the graph data~\cite{besta2020practice}. In contrast to native graph stream frameworks, like STINGER~\cite{ediger2012stinger}, LLAMA~\cite{macko2015llama}, GraphIn~\cite{sengupta2016graphin} or GraphTau~\cite{iyer2016time}, the system Graphflow~\cite{kankanamge2017graphflow} is a in-memory property graph storage that supports continuous subgraph queries.

% Bitemporal graph databases

\section{TPGM$^+$: A bitemporal property graph model}
\label{sec:tpgm}

A property graph is characterized by vertices (or nodes) and directed edges, e.g., users in a social network and their friendship relationships. The graph is labeled, which means, according to the definition, nodes and edges can be assigned one or more type labels, e.g. \textit{User} or \textit{Friendship}. Nodes and edges can have zero or more so-called properties, which are represented as key-value pairs and describe the entity or relationship in more detail, e.g., \textit{name:Christopher} or \textit{city:Leipzig}. By default, a property graph is also a multigraph, i.e., it is permitted to have multiple edges (also called parallel edges) between two nodes.

Based on the PGM, we previously introduced the Temporal Property Graph Model (TPGM)~\cite{rost2019analyzing} and formally defined it in~\cite{rost2021distributed}. It extends the PGM mainly by adding two additional attributes to each vertex and edge that describe its validity (also denoted as lifetime) in a bitemporal way. One attribute defines the valid-time (also called application time) which describes the validity of the entity or relationship in the real world, e.g., when a call between two users starts and ends. The other attribute defines the transaction-time (also called system-time) which describes when the information about the existence of the entity or relationship was inserted into the database. This bitemporal modeling is already used in relational databases~\cite{kulkarni2012temporal}. 

One downside of the TPGM is the weak support for property changes. If a property of a vertex or edge changes, i. e., it is created updated or removed, a logical copy of this element is created that stores the updated state and the information about the time of the update. All properties that are not affected by the changes stay unchanged and exist as overhead. High frequent changes of properties and their values thus result in a huge amount of duplicated elements. 
%Danica: i think counterpart should be replaced with drawback or downside in the first sentence% -> done.

To overcome this, we created in this collaboration the $TPGM^+$, an extended version of the TPGM that adds bitemporal modeling to the layer of properties. For each $TPGM^+$ graph, two linerary ordered discrete time domains $\Omega$ exist: $\Omega^{tx}$ for the transaction time and $\Omega^{val}$ for the valid-time. Each instant in time is a time point $t_i$. The linear ordering is described as $t_i<t_{i+1}$, i.e., $t_i$ happened before $t_{i+1}$. Per time domain, each vertex, edge, \textbf{and property value} has an associated time-period $\tau$. Given $t_s,t_e\in\Omega$, a time-period $\tau=[t_s,t_e)$ is defined as a close-open time interval starting at and including $t_s$ and ending at but excluding $t_e$. The time points of $\tau$ are thus a set $\{t \mid t\in\Omega \land t_s\leq t < t_e \}$. The length or size of a period $\tau$ is defined by $l(\tau)$. Default values for lower and upper bounds are $t_{min}=-\infty$ and $t_{max}=\infty$.

%\begin{figure}
%    \centering
%    \includegraphics[width=1\textwidth]{images/example-tpgm-graph.pdf}
%    \caption{Example of a TPGM$^+$ graph with updates of vertex properties.}
%    \label{fig:tpgm+-example}
%\end{figure}

%Figure~\ref{fig:tpgm+-example} shows an example of a $TPGM^+$ graph that demonstrates the evolution of a property value.

Further, we defined several integrity constraints of a $TPGM^+$ graph, to ensure the consistency of the graph at each point in time. Each constraint is valid per time domain.
\begin{enumerate}
    \item \textit{Unique vertices, edges and properties}. At every point in time, vertices, edges and properties are unique, i.e., exist at most once. The uniqueness constraint of vertices and edges is a combined key of the element's identifier and the end-timestamp. The uniqueness constraint of a property is its name.
    \item \textit{Referential integrity of edges.} For each edge, the time intervals associated with its source and target vertices must contain the edge’s time interval. In other words, an edge can only exist when its incident vertices exist.
    \item \textit{Referential integrity of properties.} For a property value, the interval of the vertex/edge must contain the interval of the property value. In other words, a property value can only exist when the corresponding vertex/edge exists.
    \item \textit{Constant edges.} Source and target vertices of an edge never change at existence. 
    \item \textit{Constant types.} Vertex and edge labels, as well as property key names, never change.
\end{enumerate}

\section{T-PGQL: A temporal graph query language}
\label{sec:tpgql}
We will now have a look at some preliminaries needed to understand the extensions we made to the graph query language PGQL, that are explained afterwards. 
%Danica: I don't think you need a reference here since you're referencing the same section% -> done.

\subsection{Preliminary: PGQL - A (Non-temporal) Graph Query Language}\label{sec:pgql-origin}

PGQL~\cite{van2016pgql,pgql} combines graph pattern matching with SQL-like syntax and functionality and has full-blown support for regular path queries and graph construction. Because its syntax is SQL-like, the language is intuitive to use for existing SQL users. Furthermore, PGQL queries return a “resultset” with variables and bindings, just like in SQL. In this work, we refer to PGQL version 1.3. 

Matching graph patterns is one main functionality of PGQL realized by a SELECT query. Similar to SQL, a SELECT query is composed of several clauses, starting with the mandatory SELECT clause and FROM clause. In PGQL, the SELECT clause defines the returned result entities of the query. Since the result of a SELECT query is always a table, the SELECT clause defines the attributes of the result table. The following syntactic structure of a PGQL SELECT query and clause is taken from~\cite{pgql}:

\begin{verbatim}
SelectQuery    ::=  SelectClause
                    FromClause
                    WhereClause?
                    ...
SelectClause    ::= 'SELECT' 'DISTINCT'? ExpAsVar ( ',' ExpAsVar )*
                  | 'SELECT' '*'
ExpAsVar    ::= ValueExpression ( 'AS' VariableName )?
\end{verbatim}

The graph pattern to be matched is defined by the FROM clause that includes one or multiple MATCH clauses. A MATCH clause defines a path- or graph pattern, where a graph pattern is a composition of path patterns.

The following syntactical definitions of the FROM- and MATCH clauses are taken from~\cite{pgql}. 
\begin{verbatim}
FromClause    ::= 'FROM' MatchClause ( ',' MatchClause )*
MatchClause   ::= 'MATCH' ( PathPattern | GraphPattern ) OnClause?
GraphPattern  ::= '(' PathPattern ( ',' PathPattern )* ')'
PathPattern   ::=   SimplePathPattern | ShortestPathPattern | ...
\end{verbatim}

Path patterns describe topology constraints, where a topology constraint is a composition of one or multiple vertices and edges. A vertex or edge is optionally identified by a variable, i.e., a symbolic name to reference it in other clauses. It is also possible to define one or more label predicates directly in the pattern. A PGQL SELECT query returning all person and movie names that match this pattern can be formulated as follows:

\begin{Verbatim}[numbers=left]
SELECT p.name, m.movie
FROM MATCH (p:Person)-[l:like|dislike]->(m:Movie)
\end{Verbatim}

For further information, we refer to the official documentation of PGQL, available at~\cite{pgql}.

\subsection{Extensions of PGQL to Query TPGM$^+$ Graphs}
In current graph query languages patterns (or paths) are searched in the whole available graph database without observance of any evolution. Having a temporal graph modeled by the TPGM$^+$ (see Sect.~\ref{sec:tpgm}), several new requirements arise for a query language to support temporal features of the model. In this work, we limit to the retrieval of data. The manipulation of data as well as functions for the definition and manipulation of database structures are not considered yet and part of future work. 

\subsubsection{Access of Temporal Attributes}\label{sec:projection-selection}
Our data model tracks changes in a bitemporal model on a level of vertices, edges and properties. We introduce the possibility to add these attributes via projection to the resulting relation and to use these attributes in expressions for selection.

We distinguish four different temporal characteristics of a vertex, edge and property: the period itself, its lower bound timestamp (inclusive), its upper bound timestamp (exclusive) and length/duration of a period. 

The identifier for the period is defined as \texttt{VAL\_TIME} for the valid time domain and \texttt{TX\_TIME} for the transaction time domain. A period is, like in SQL, not a data type but a type definition~\cite{kulkarni2012temporal}. The textual representation a period projection is a concatenated result of both period bounds in the form of: \texttt{[\{from\},\{to\})}. A period type can be used for several predicates, as later described. The identifiers for the period bounds are \texttt{VAL\_FROM} and \texttt{VAL\_TO} for the valid time domain and \texttt{TX\_FROM} and \texttt{TX\_TO} for the transaction time domain. The result of a period-bound access is a timestamp.

To access these attributes of a vertex or edge, they can be used like a property access by dot notation, e.g., \texttt{var1.TX\_FROM}, \texttt{var2.VAL\_TIME}. According to the temporal attributes of a property, the same suffix can be used on a property access expression, like \texttt{var1.prop1.TX\_FROM} or \texttt{var2.prop2.VAL\_TIME}. The following syntax describes the notation.

\begin{Verbatim}
TimeIdentifier  ::= 'TX_TIME' | 'VAL_TIME' | 'TX_FROM' | 'TX_TO' 
                  | 'VAL_FROM' | 'VAL_TO'
Property        ::= Identifier
VarRef          ::= Identifier
PropRef         ::= VarRef '.' Property
ElementTimeRef  ::= VarRef '.' TimeIdentifier
PropertyTimeRef ::= PropRef '.' TimeIdentifier
TimeRef         ::= ElementTimeRef | PropertyTimeRef
\end{Verbatim}

For example, the following query returns all available temporal characteristics of person vertices and their name property.

\begin{Verbatim}[numbers=left]
SELECT n.TX_FROM, n.TX_TO, n.TX_TIME, 
       n.VAL_FROM, n.VAL_TO, n.VAL_TIME, 
       n.name.TX_FROM, n.name.TX_TO, n.name.TX_TIME, 
       n.name.VAL_FROM, n.name.VAL_TO, n.name.VAL_TIME
FROM MATCH (n:Person) ON student_network
\end{Verbatim}

Besides the period of a graph element or its property, the length/duration of this period can be queried. We introduce a \texttt{LENGTH([unit,]period)} expression which consumes a period access identifier (\texttt{period}) as argument and an optional unit (i.e., \texttt{YEAR, QUARTER, MONTH, WEEK, DAY}, \texttt{HOUR}, \texttt{MINUTE}, \texttt{SECOND}, \texttt{MILLISECOND}). If no unit is given, the default unit \texttt{MILLISECOND} is used. This expression returns a numerical value that can be used within several expressions, e.g., binary constraints. The following query returns the duration in days of the valid time period of a person's name property for all persons whose valid time exceeds one day.

\begin{Verbatim}[numbers=left]
SELECT LENGTH(DAY, n.name.VAL_TIME)
FROM MATCH (n:Person) ON student_network
WHERE LENGTH(DAY, n.name.VAL_TIME) > 1
\end{Verbatim}

\subsubsection{Temporal Filtering and Chronological Pattern Matching}

Graphs with a managed valid-time are intended for meeting the requirements of applications that  are  interested in capturing  time  periods  during  which  the  data  is (believed to be) valid in the real world. For each vertex or edge, as well as their properties, a valid-time period is available. As described above, the identifier of this period is \texttt{VALID_TIME} and the beginning and ending bounds are \texttt{VAL_FROM} and \texttt{VAL_TO}. They can be used as a suffix to variable and property access. Analogous to the transaction time attributes, the valid time period returns a new \textit{period} type definition and the bounds return a timestamp type. Latter can be used like regular timestamp attributes, e.g., within binary relations in the WHERE clause. The extended PGQL WHERE clause~\cite{pgql} is defined as follows\footnote{The \texttt{TemporalExpression} is a temporal extension of our work.}:

\newpage
\begin{verbatim}
WhereClause        ::= 'WHERE' ValueExpression
ValueExpression    ::= VariableReference
                     | PropertyAccess
                     | ...
                     | TemporalExpression
TemporalExpression ::= ElementTimeRef
                     | PropertyTimeRef
                     | Overlaps | Equals
                     | Precedes | Succeeds
                     | Contains
Overlaps           ::= TimeRef 'OVERLAPS' TimeRef
Equals             ::= TimeRef 'EQUALS' TimeRef
Precedes           ::= TimeRef ('IMMEDIATELY')? 'PRECEDES' TimeRef
Succeeds           ::= TimeRef ('IMMEDIATELY')? 'SUCCEEDS' TimeRef
Contains           ::= TimeRef 'CONTAINS' TimeRef
\end{verbatim}

For example, to retrieve all students who studied at a University in Leipzig as of February 15, 2019, one can express the query by accessing the period bounds in predicates of the WHERE clause:

\begin{Verbatim}[numbers=left]
SELECT n.name
FROM MATCH (n:Person)-[s:studiedAt]->(u:University)
WHERE u.city = 'Leipzig'
    AND s.VAL_FROM <= DATE '2019-02-15'
    AND s.VAL_TO > DATE '2019-02-15'
\end{Verbatim}

To simplify the expression of such predicates, several language extensions are further defined. For example, we can use one of the  period  predicates provided in SQL:2011  for  expressing  conditions  involving  periods: \texttt{CONTAINS}, \texttt{OVERLAPS}, \texttt{EQUALS}, \texttt{(IMMEDIATELY) PRECEDES}, and \texttt{(IMMEDIATELY) SUCCEEDS}. All period predicates need two expressions that return a period as arguments, except for \texttt{CONTAINS}, which also allows a timestamp as a second argument. The query above can be simplified by using the \texttt{CONTAINS} predicate:

\begin{Verbatim}[numbers=left]
SELECT n.name
FROM MATCH (n:Person)-[s:studiedAt]->(u:University)
WHERE u.city = 'Leipzig'
    AND s.VALID_TIME CONTAINS DATE '2019-02-15'
\end{Verbatim}

To retrieve all students of Universities in Leipzig who are matriculated from January  1,  2018,  to  January  1,  2019,  one could formulate the query by using a temporal condition, in our case, the \texttt{OVERLAPS} predicate:

\begin{Verbatim}[numbers=left]
SELECT n.name
FROM MATCH (n:Person)-[s:studiedAt]->(u:University)
WHERE u.city = 'Leipzig'
    AND s.VALID_TIME OVERLAPS PERIOD(DATE '2018-01-01', 
                                     DATE '2019-01-01')
\end{Verbatim}

In the example, we also show a new period constructor expression that allows the creation of a period instance from two timestamps $t_1$ and $t_2$ with $t_1 <= t_2$. The timestamps can be defined by any expression that returns a date or timestamp instance. For example, they can be created through a date or timestamp constructor as in the example or extracted from a graph element or property through a period bound identifier, e.g., \texttt{PERIOD(DATE '2018-01-01', x.VAL_TO)}.

Also, the transaction time period of elements and properties can be used in predicates of this kind. The following query returns the name and system-time period for students matriculated in a University located in Leipzig where the information about the matriculation was added to the database after March 1, 2018. The syntax part \texttt{FOR TX_TIME ALL} is explained in the next section.

\begin{Verbatim}[numbers=left]
SELECT n.name, n.TX_TIME
FROM MATCH (n:Person)-[s:studiedAt]->(u:University) FOR TX_TIME ALL
WHERE u.city = 'Leipzig'
    AND s.TX_FROM >= TIMESTAMP '2018-03-01 00:00:00'
\end{Verbatim}

The classical graph pattern matching is used to find a matching subgraph in the graph database that matches exactly with the defined query pattern. In the well-known static scenario, a query pattern has no information about the chronological ordering of the given entities and relationships. For example, a pattern like \texttt{(p1:Person)-[l1:likes]->(p2:Person)-[l2:likes]->(p3:Person)} includes no information about when the likes happened or if one like happened before the other of if they happened at the same time. To overcome this lack, the above introduced temporal predicates can be used to enrich such patterns with temporal information. The following code exemplifies that:

\begin{Verbatim}[numbers=left]
SELECT p1.name, p2.name, p3.name
FROM MATCH (p1:Person)-[l1:likes]->(p2:Person)-[l2:likes]->(p3:Person)
WHERE l1.VAL_TIME PRECEDES l2.VAL_TIME
\end{Verbatim}

Here we add a constraint, that the like between p1 and p2 must happen before the like between p2 and p3. These kinds of predicates can thus be used to define a temporal ordering in a path pattern.

\subsubsection{Graph Snapshot Retrieval and Historical Pattern Matching}\label{sec:from}

T-PGQL can be used to find matching subgraphs in a specific state of the temporal graph with respect to the transaction time domain. This state can be a snapshot at a defined timestamp or all changes of a given period. Former represents a valid snapshot graph without multiple versions of a single instance. The transaction time dimension is reduced to a single point in time. Latter is again a temporal property graph that can have multiple versions of a single instance. The transaction time dimension is reduced to a range in time.

To search for a defined pattern using this kind of time traveling, we extended PGQL's FROM clause with a clause similar to the SQL extension for temporal tables. The name of the transaction time period definition is fixed to \texttt{TX_TIME}. To query the historical data, the clause \texttt{FOR TX_TIME \{predicate\}} has to be used directly after a MATCH clause (see~\Cref{sec:pgql-origin}). To define the timestamp or period to query for, we provide four predicates as syntactic extensions:

\begin{Verbatim}
GraphMatch          ::= 'MATCH' PathPattern OnClause? SysTimeCond
SysTimeCond         ::= 'FOR' 'TX_TIME' ( AsOf | FromTo 
                                        | BetweenAnd | 'ALL')
AsOf                ::= 'AS' 'OF' TimeRef
FromTo              ::= 'FROM' TimeRef 'TO' TimeRef
BetweenAnd          ::= 'BETWEEN' TimeRef 'AND' TimeRef
\end{Verbatim}

The argument \texttt{TimeRef} could be any expression returning a timestamp attribute, i.e., a date or timestamp constructor (e.g., \texttt{TIMESTAMP('2020-01-01')}), current timestamp expression (\texttt{CURRENT_TIMESTAMP}) or access expressions of temporal attributes for vertices, edges or properties. If the \texttt{FOR TX_TIME} clause is not used, the result will show the current data, as if one had specified \texttt{FOR TX_TIME AS OF CURRENT_TIMESTAMP}. Thus, the following queries are equal:

\begin{Verbatim}[numbers=left]
SELECT n.name
FROM MATCH (n:Person) 
\end{Verbatim}

\begin{Verbatim}[numbers=left]
SELECT n.name
FROM MATCH (n:Person) FOR TX_TIME AS OF CURRENT_TIMESTAMP
\end{Verbatim}

The first predicate \texttt{AS OF \{timestamp\}} is used to see the graph as it was at a specific point in time in the presence or past. The following example query retrieves the graph as it was on 1st February 2020 at 1 pm.

\begin{Verbatim}[numbers=left]
SELECT n.name
FROM MATCH (n:Person) FOR TX_TIME AS OF TIMESTAMP '2020-02-01 13:00'
\end{Verbatim}

The next query using the \texttt{BETWEEN \{timestamp\} AND\{timestamp\}} predicate will show all graph elements that were visible at any point between two specified points in time. It works inclusively, i.e., an element visible exactly at the start or exactly at the end will be added to the result set too. 

\begin{Verbatim}[numbers=left]
SELECT n.name
FROM MATCH (n:Person) FOR TX_TIME BETWEEN 
     TIMESTAMP '2020-02-01 12:00' AND TIMESTAMP '2020-02-28 12:00:00'
\end{Verbatim}

The extension \texttt{FROM \{timestamp\} TO \{timestamp\}} will also show all elements that were visible at any point between two specified points in time, including start, but excluding end.

\begin{Verbatim}[numbers=left]
SELECT n.name
FROM MATCH (n:Person) FOR TX_TIME FROM TIMESTAMP '2020-02-01 12:00' 
    TO TIMESTAMP '2020-02-28 12:00:00'
\end{Verbatim}

To query for the current state and complete history of a given pattern the predicate \texttt{ALL} can be used.

\begin{Verbatim}[numbers=left]
SELECT n.name
FROM MATCH (n:Person) FOR TX_TIME ALL
\end{Verbatim}

In PGQL, it is possible to define multiple patterns within a single FROM clause by using multiple match clauses separated by a comma. Our extension can be used within each of these match expressions. This provides a flexible mechanism to define patterns with parts occurring at different times. For example, to find people who currently liked a post that already existed on January 1st, 2020, the following query can be used:

\begin{Verbatim}[numbers=left]
SELECT m.firstName, m.lastName
FROM MATCH (p:Post) FOR TX_TIME AS OF DATE ‘2020-01-01’,
     MATCH (m:Person)-[:likes]->(p) // current
\end{Verbatim}

Another query scenario is to find a sensor that is currently (\texttt{FOR TX_TIME AS OF CURRENT\_TIMESTAMP}) connected to an asset that existed in the past (\texttt{FOR TX_TIME AS OF DATE '...'}). Besides a path pattern, a graph pattern is a concatenated list of path patterns in round brackets, which can be also specified after the match keyword. If a transaction time predicate should be applied to a set of path patterns, it can be used after such a graph pattern, as can be seen in the following example.

\begin{Verbatim}[numbers=left]
SELECT m.firstName, m.lastName
FROM MATCH (
    (p:Post)-[:hasTag]->(t:Tag)-[:inClass]->(tc:TagClass),
    (m:Person)-[:likes]->(p:Post) 
    ) FOR TX_TIME AS OF DATE ‘2020-01-01’   
\end{Verbatim}

\subsubsection{Bitemporal Queries}

A TPGM graph has both a managed transaction- and valid-time domain. Graph elements, as well as their properties, are associated with both transaction-time and valid-time periods. This concept is very useful for use cases where it is necessary to capture both the periods during which facts were believed to be true in real-world as well as periods during which those facts were recorded in the database.

For example, a student changes his address. Typically the  address changes legally at a specific time, but it is not changed in the database concurrently with the legal change. In that case, the transaction-time period automatically records when the new address is known to the database, and the valid-time  period  records  when  the  address  was  legally  effective. Successive updates to bitemporal graphs can journal complex twists and turns in the state of knowledge captured by the database~\cite{kulkarni2012temporal}.

Queries on bitemporal graphs can specify predicates on both dimensions to qualify rows that will be returned as the query result.  For example, the following query returns all students of Universities in Leipzig who are matriculated as of December 1, 2019, recorded in the graph database as of January 1, 2020.

\begin{Verbatim}[numbers=left]
SELECT n.name
FROM MATCH (n:Person)-[s:studiedAt]->(u:University) 
    ON student_network FOR TX_TIME AS OF DATE '2020-01-01'
WHERE u.city = 'Leipzig'
    AND s.VALID_TIME CONTAINS DATE '2019-12-01'
\end{Verbatim}

\subsubsection{Query the Evolution of a Property}
Every vertex and edge can have zero, one or more properties in form of key-value pairs, where the key represents the name of the property. For every property, transaction-time versioning is supported to track the addition of new properties, changes in values or deletion of exiting properties. Thus a property behaves like a vertex or an edge. By inserting a vertex or edge with properties into the database, each gets the same system-time period as the respective vertex or edge.

The previous introduced FROM clause extensions have also an effect on the property retrieval since their transaction-time period will be considered, too.

In addition, each property contains a valid-time period. For example, each \texttt{University} vertex has a property \texttt{studentCount} whose value is periodically updated. Each value has an application time period that defines, for which time the value was true. If no further condition is specified for an application time enabled property, all values are returned.

\begin{Verbatim}[numbers=left]
SELECT u.name as name, u.studentCount as cnt
FROM MATCH (u:University)
WHERE u.city = 'Leipzig'
\end{Verbatim}

Result:
\begin{Verbatim}
+---------------------+-------+
|        name         |  cnt  |
+---------------------+-------+
| Leipzig University  | 28004 |
| Leipzig University  | 28797 |
| Leipzig University  | 29061 |
+---------------------+-------+
\end{Verbatim}

To retrieve the information of the validity of the values, the period of validity can be selected. %Danica: "its period" is a big vague to me, consider expanding to make it more clear, e.g by saying "the period of validity can be selected"% -> done.

\begin{Verbatim}[numbers=left]
SELECT u.name as name, u.studentCount as cnt, 
    u.studentCount.VALID_TIME as validity
FROM MATCH (u:University)
WHERE u.city = 'Leipzig'
\end{Verbatim}
\newpage
Result:

\begin{Verbatim}
+---------------------+-------+--------------------------------------+
|        name         |  cnt  |                  validity            |
+---------------------+-------+--------------------------------------+
| Leipzig University  | 28004 | [2016-04-01 00:00, 2017-04-01 00:00) |
| Leipzig University  | 28797 | [2017-04-01 00:00, 2018-04-01 00:00) |
| Leipzig University  | 29061 | [2018-04-01 00:00, 2019-04-01 00:00) |
+---------------------+-------+--------------------------------------+
\end{Verbatim}

All previously introduced conditions of the WHERE clause that are applicable for vertices and edges can be used by properties too.

\begin{Verbatim}[numbers=left]
SELECT u.name, u.studentCount, u.studentCount.VALID_TIME as validity
FROM MATCH (u:University)
WHERE u.city = 'Leipzig'
    AND u.studentCount.VALID_TIME CONTAINS TIMESTAMP '2018-01-01 00:00'
\end{Verbatim}
Result:
\begin{Verbatim}
+---------------------+-------+--------------------------------------+
|        name         |  cnt  |                  validity            |
+---------------------+-------+--------------------------------------+
| Leipzig University  | 28797 | [2017-04-01 00:00, 2018-04-01 00:00) |
+---------------------+-------+--------------------------------------+
\end{Verbatim}

\subsection{Aggregations}
An aggregate function allows performing a calculation on a set of values to return a single scalar value. Aggregate functions are used with the GROUP BY and HAVING clauses of the query.

In this work, the PGQL language was extended by two functions: \texttt{FIRST({date or timestamp values})} and \texttt{LAST({date or timestamp values})}. The former returns the chronological earliest date or timestamp, while the latter returns the chronological last. 

The following query returns the first beginning of a \textit{studentAt} relationship according to the application-time and system-time domain.

\begin{Verbatim}[numbers=left]
SELECT FIRST(s.VAL_FROM) as earliestStart, 
       FIRST(s.startTT) as earliestTx
FROM MATCH ()-[s:studiedAt]->()
\end{Verbatim}

Result:

\begin{Verbatim}
+---------------------+---------------------+
|    earliestStart    |     earliestTx      |
+---------------------+---------------------+
| 1409-04-01 00:00:00 | 2006-05-12 14:45:22 |
+---------------------+---------------------+
\end{Verbatim}

The following query can be used to answer the question: For universities of a certain city, when was the first time a student began his studies, and when was the most recent time? %Danica: consider replacing latest to "most recent" to make it more direct% -> done.

\begin{Verbatim}[numbers=left]
SELECT u.city, 
       FIRST(s.VAL_FROM) as earliestStart, 
       LAST(s.VAL_FROM) as latestStart
FROM MATCH (n:Person)-[s:studiedAt]->(u:University)
GROUP BY u.city
\end{Verbatim}

Result:

\begin{Verbatim}
+---------+---------------------+---------------------+
|  city   |    earliestStart    |     latestStart     |
+---------+---------------------+---------------------+
| Leipzig | 1409-04-01 00:00:00 | 2020-04-01 00:00:00 |
| Berlin  | 1810-04-01 00:00:00 | 2020-04-01 00:00:00 |
| Munich  | 1472-04-01 00:00:00 | 2020-04-01 00:00:00 |
+---------+---------------------+---------------------+
\end{Verbatim}

\section{CGN: Event detection through Continuous Graph Notifications}
\label{sec:cgn}

We introduced T-PGQL as a query language for executing SELECT queries on TPGM$^+$ graphs in the previous Section~\ref{sec:tpgql}. Querying a graph with T-PGQL is usually done in a single graph query execution, i.e., a user formulates a SELECT query, executes that query on the current state of a database system that maintains a TPGM$^+$ graph and gets a result in form of a table back. In this way, one can query for current and historical data of the graph where one query leads to one fixed result. Subsequent changes of the graph are not taken into account unless the query is executed again which leads to a result that recognizes all transactions that are made until the time when the query was executed.

Talking about changes in a graph leads to the observation of events. An event itself can be everything that happened at a defined point in time, e. g., an asset's sensor captures a temperature or a user liked a post in a social network. We semantically distinguish between \textit{application-world events} and \textit{transaction-world events}. Former is an event that happened in the observed real world. It can be described by a graph pattern and its predicates inside a graph query. In contrast, the discovery of an instance of a pattern in the data store at a specific time is a \textit{transaction-world event}. I. e., the most recent commit (1) created an instance of a pattern that did not exist before, e. g., a captured temperature of a sensor exceeds a threshold, or (2) destroys an instance of the pattern that already existed, e.g. a friendship relation between two users of a social network is removed. 

The question arises, how a user can get a notification about a \textit{transaction-world events}, i. e. when graph data changes that affect the elements that are accessed to create the query result. For RDBMS, there is a feature called Continuous Query Notification (CQN)~\cite{cqn-oracle-docs} which is currently implemented by the Oracle database. It allows to register a SQL query and receive notifications when an event occurs that changes a table, i. e., that rows have been updated. This is useful in applications like near real-time monitoring, auditing applications, or for such purposes as mid-tier cache invalidation~\cite{cqn-oracle-docs}. In this work, we adapt some concepts of the relational CQN to the graph domain. On a graph database that implements the TPGM$^+$, it is not only possible to simply execute a T-PGQL query, but also to register one. A registration is configured by the T-PGQL SELECT query itself, a registration validity period that specifies when the notification is enabled and when it will be disabled, a notification endpoint (e.g., a messaging queue) and a notification type. There are two types of graph query notifications: (1) the \textit{Graph Change Notification (GCN)} and (2) the \textit{Graph Query Result Change Notification (GRCN)}~\cite{zimmer2021kontinuierliche}. The GCN notifies the registrar if a transaction on the queried graph affects the graph elements queried. This does not automatically mean that the query result has also changed. To get a notification about an event that changes the graph in a way that affects the query result, the GRCN type must be used. Note that both types are a smart evaluation of a registered query, but not a query re-execution. This means the recipients are notified about the event but do not get the updated query result. 

Assume the following T-PGQL query. The query describes the event of a temperature measurement above a value of 40 from a sensor that is part of an asset that is connected to an asset with id '42'. The projection of the query is the sensor value and its validity timestamp, which describes the time when the measure happened in the observed real world. 

\begin{Verbatim}[numbers=left]
SELECT s.value, s.value.VAL_FROM
FROM MATCH (a1:Asset)-[p:partOf]->(a2:Asset)-[:hasSensor]->(s:Sensor)
WHERE a1.id = 42 AND s.type = 'temperature' AND s.value > 40
\end{Verbatim}

We assume that there are several matches for this pattern without recognizing the predicate of the value threshold, but none that fulfills this condition (i.e., all temperature values are below a value of 40). Further assume, that at a time $t_1$, the value of a sensor's property is updated from 39 to 40. If the query is registered by a GCN, a notification about that event is created, since one of the involved graph elements changed its state, but the query result is not affected. If the query is registered by a GRCN, no notification is created, since the query result does not change (it is still empty). Now, assume that at time $t_2$ the value of the sensor changes from 40 to 41. At this time, a notification is created for both types, since (1) a graph element that is part of the pattern changes in some way and (2) the query result changes in the form that now one row is part of the result.

\section{BiTeGra: A bitemporal graph management system}
\label{sec:bitegra}
With the development of a bitemporal graph model, a query language for bitemporal graphs and the possibility of registering queries for continuous evaluation, the foundations for a graph database have been laid that unites all of the concepts considered. We will introduce an architectural draft of BiTeGra, a bitemporal graph database management system, and details of three main components including the graph to relation mapping, the graph modification features and the querying possibilities. Figure~\ref{fig:architecture} shows an architectural draft of the BiTeGra system. The core of the system is a relational database management system with bitemporal table support. The graph data is stored in this database in a way that is described in Section~\ref{sec:schema}.

\subsection{Graph Modifications}
A graph modification API provides an interface to manipulate graph data. There are methods to allow the following modifications to the maintained temporal graph:

\begin{figure}[t]
    \centering
    \includegraphics[clip, trim=0 6cm 0 0, width=1.00\textwidth]{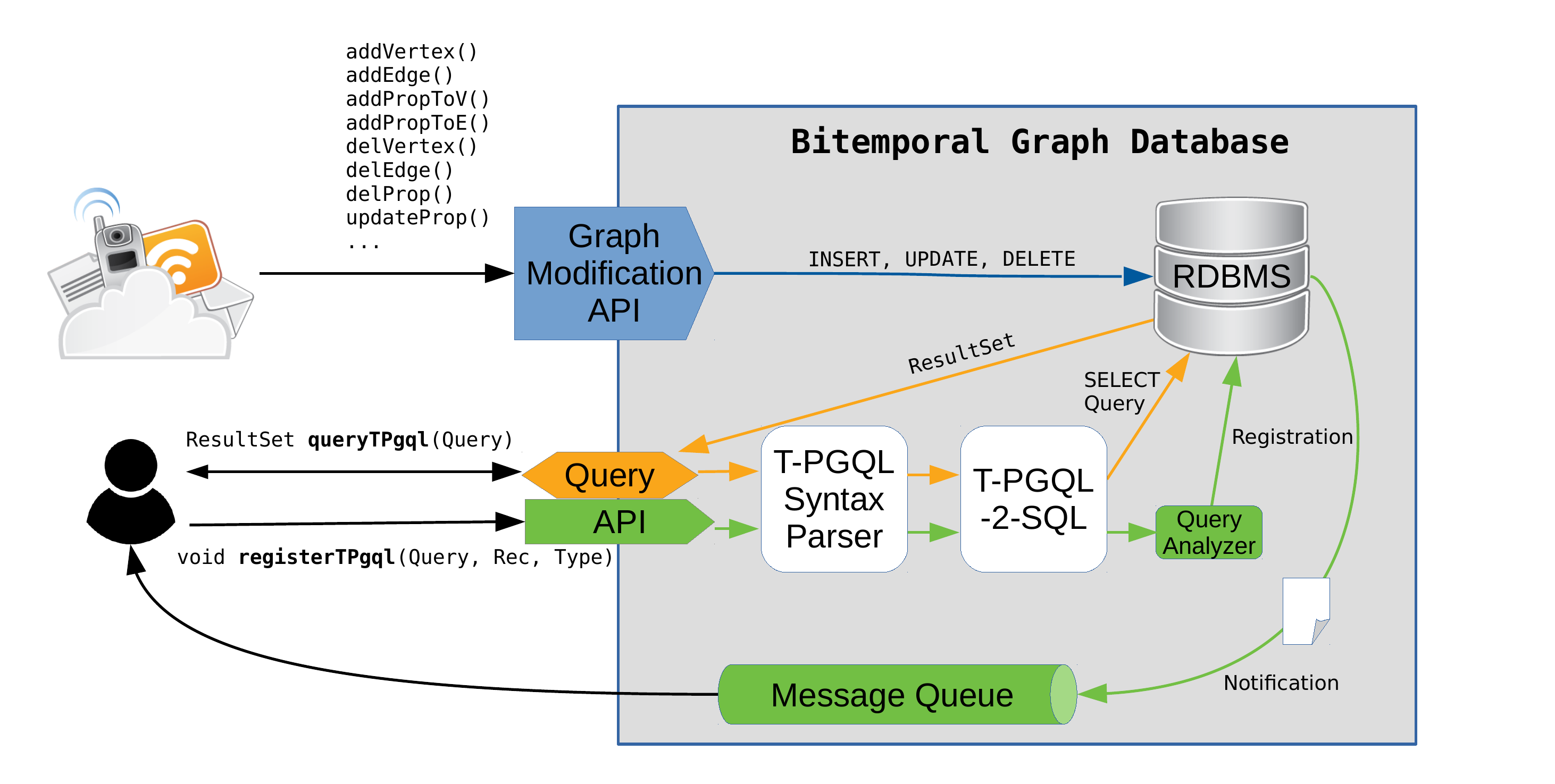}
    \caption{Architectural draft of the bitemporal graph database}
    \label{fig:architecture}
\end{figure}

\begin{itemize}
    \item Add vertices/edges
    \item Add properties to vertices/edges
    \item Edit properties
    \item Edit valid-time attributes of vertices/edges
    \item Edit valid-time attributes of properties
    \item Delete properties from vertices/edges
    \item Delete vertices/edges
\end{itemize}

Each API call is internally translated to a SQL INSERT, UPDATE or DELETE query, depending on the used table schema. The API can be used by a wide variety of applications, e.g., the import of graph streams and other streaming and time-series data.

\subsection{Schema mapping: Graph to Table and vice versa}
\label{sec:schema}

One important design choice when storing a graph structure in a relational system is the table schema that is used. It impacts the complexity of translated queries and thus the runtime for fetching a result. Fritzsche has evaluated different approaches in~\cite{fritzsche2021relationale}, which are partly demonstrated in the following.

\paragraph{\textbf{Table schema for vertices and edges}} The first types of entities that need to be considered are the vertices and edges themselves. In this work, we compared two commonly used schemata to represent vertex and edge datasets as relations. 

The first one is the \textbf{GVE-Schema} (\textbf{G}raph \textbf{V}ertex \textbf{E}dge-Schema) or \textbf{Vertical Schema}~\cite{adameit2020evaluation,saalmann2019relationale,rost2021distributed}. It mainly uses two tables - one for vertices and one for edges. Each table has an identifier and a label attribute. The table used for edges has, in addition, two attributes storing the source and target vertex identifier of an edge. To represent bitemporal graphs, two additional attributes represent the lower and upper bound of the valid time period whereas two additional attributes represent the lower and upper bound of the transaction time period. The representation of properties is discussed later. To ensure the integrity of the graph, primary and foreign key constraints are used. For vertices and edges, the primary key is a combination of the identifier and the transaction period ending timestamp. The edge table has, in addition, two foreign keys pointing to the vertex table - one for the source vertex identifier and one for the target vertex identifier. One advantage of this schema is that it is flexible about new labels since labels are an attribute of the table. Another advantage is that querying for all vertices or edges can be done in one step. A disadvantage is the size of the two tables, which can be large. 
Also joining this kind of large tables might be an issue in terms of performance.

The second one is the \textbf{TFL-Schema} (Tables For Labels-Schema) or \textbf{Horizontal Schema}~\cite{adameit2020evaluation,saalmann2019relationale}. Here, vertices and edges are separated by their label and stored in one table per label.
The specialty of the edge tables is that the node labels of the source and target nodes are also taken into account when separating. For example, there is one table of \textit{like} edges between vertices of type \textit{Person} and one other table of \textit{like} edges that connect vertices of type \textit{Person} and \textit{Post}. This is because the foreign keys point to potentially different tables storing source and target vertices.
The advantage of the TFL-Schema is the reduced table size compared to the GVE-Schema. This is particularly conceivable for answering path queries that needs joining tables. Another advantage is that queries for a specific node and edge type can 
simply query for the entire table, whereas filtering is required in the GVE-Schema.
A disadvantage of this schema is that all tables of nodes or edges have to be queried if no label is specified in a query. Adding nodes or edges with new labels to the graph also may require additional tables to store the corresponding element types.

To overcome the limitations of both schema models and combine their advantages, we introduce the \textbf{HyVE-Schema} (\textbf{Hy}brid Schema for \textbf{V}ertices and \textbf{E}dges). As the name suggests, this is a hybrid of the two schemes mentioned and leaves a decision open depending on the application. For each vertex and edge label, it can be decided whether the elements will be stored in a general vertex or edge table, or in a label-specific table. A metadata table stores these decisions and can be queried if the information about the location of a specific vertex and edge type is needed.

\paragraph{\textbf{Table schema of properties}} Another important design decision is the modeling of properties. Again, we differentiate two ways of representing properties of vertices and edges in the relational schema.

The first one is \textbf{PAC} (\textbf{P}roperties \textbf{a}s \textbf{C}olumns), which, like the name suggests, stores properties as additional columns in the vertex and edge tables (independent of the schema for vertices and edges). The column or attribute name is defines the property key whereas the values in the entities correspond to property values. The bitemporal modeling is implicitly used from the tables containing the columns.
One advantage of this approach is the usage of datatypes for property values that are supported by SQL. For example, a property \texttt{name} can be represented by an attribute with the same name and datatype \texttt{VARCHAR}. This way, a datatype may be assigned to each property. Another advantage is that no joins need to be executed when accessing the properties. If an element does not have a property, the respective column value is set to null. Disadvantages are thus many null values for elements that do not have a property that is modeled as a column. Another disadvantage is the need for change of the table schema if new property types are added, which results in the addition of a new column. The last weak point of this approach is that the whole table entry for a vertex or edge has to be copied when system-versioning is enabled. Even if only a single property value changes, a copy of the whole row is created which gets the new value and thus leads to many redundant attribute values.

The second approach is \textbf{PAT} (\textbf{P}roperties \textbf{a}s \textbf{T}able). Here, a property table per vertex or edge table stores all properties. A property table has mainly 3 attributes: the identifier of the parent element (vertex or edge), the key and the value. The bitemporal modeling is given by four additional attributes, as described in the GVE-Schema. A foreign key points from the element identifier to the identifier of the respective element table. One advantage is that all properties are optional, as the property graph model defines. It is easy to add new properties to a vertex or edge type just by inserting a new row in the property table. Another advantage is that the change of a property does not affect the vertex or edge entity itself, but just creates a copy in the property table for this property through the system-versioning. Disadvantages are the need for one join per property access and the fixed data type for the value column, implying that all property values must have the same datatype.

Again, to combine the strengths of both strategies and to overcome some weaknesses, we introduce the \textbf{HyPe-Schema} for bitemporal modeled vertex and edge properties. For each vertex and edge type, it can be chosen whether a property is stored as a column in the vertex or edge table (see PAC), or in a separate property table for this vertex or edge type. A metadata table stores these decisions and can be queried, if the information about the location of a specific property is needed.

\subsection{Graph Querying} 
To query a TPGM$^+$ graph that is stored in the bitemporal graph storage with T-PGQL, there are two types of queries: (1) a single query that is executed once and returns the result set back to the user (colored orange in Fig.~\ref{fig:architecture}) and (2) a query registration for continuous event notification (colored green in Fig.~\ref{fig:architecture}).

For both cases, the query string will be first parsed by a \textit{T-PGQL Syntax Parser}, which verifies and analyses the query and creates an object representation containing lists of projections, predicates, requested vertices and edges. The parser is based on the Open-Source project \textit{PGQL Parser and Static Query Validator} which is available on GitHub\footnote{\url{https://github.com/oracle/pgql-lang}}. We extended this implementation by supporting all additional temporal features of the T-PGQL language. The resulting parsed query object is then used as input for the \textit{T-PGQL-2-SQL translator}. This component creates a SQL SELECT query from the content of the query object in a way that the result of the query forms the T-PGQL result. 

For the single query execution, the SQL SELECT query is executed on the relational database and the resultset is transferred back to the user that initially called the query method. For the continuous event notification, the SQL SELECT query, which is again a representation of a transaction-world event, is given to a \textit{Query Analyser} that extracts necessary information of the query for the registration routine, e.g., touched tables, needed attributes and predicates~\cite{zimmer2021kontinuierliche}. Depending on the type of notification (GCN or GRCN), several triggers were registered on the concerned tables that create the notifications, which are sent to a message queue. The user, who has access to this queue, now receives notifications about events that either affect the related tables (GCN) or the query result (GRCN).

\section{Summary}
\label{sec:summary}

We presented a summary of our cooperation outcomes. We introduced (1) the bitemporal property graph model \textit{TPGM$^+$} which supports the evolution of property values, (2) the declarative graph query language \textit{T-PGQL} to match patterns in a TPGM$^+$ graph, (3) an event detection engine that allows the registration of T-PGQL queries on a bitemporal graph storage for continuous notifications of graph changes and (4) a prototype called \textit{BiTeGra} which stores TPGM$^+$ graphs in temporal relational tables and offers features to modify and query the graph as well as registering queries for event detection.

%
% ---- Bibliography ----
%
% BibTeX users should specify bibliography style 'splncs04'.
% References will then be sorted and formatted in the correct style.
%
\bibliographystyle{splncs04}
\bibliography{bibliography}

\end{document}